\documentclass{aa}
\usepackage{graphicx}
\usepackage[varg]{txfonts}
\usepackage{cases}
\usepackage{natbib}
\usepackage{amsmath}

\begin{document}

   \title{A revised cone model and its application to non-radial prominence eruptions}

   \author{Q. M. Zhang\inst{1}}

   \institute{Key Laboratory of Dark Matter and Space Astronomy, Purple Mountain Observatory, CAS, Nanjing 210023, PR China \\
              \email{zhangqm@pmo.ac.cn}
              }

   \date{Received; accepted}
    \titlerunning{A revised cone model}
    \authorrunning{Q. M. Zhang}

 \abstract
   {The traditional cone models achieve great success in studying the geometrical and kinematic properties of halo coronal mass ejections (CMEs).}
   {In this paper, a revised cone model is proposed to investigate the properties of CMEs as a result of non-radial prominence eruptions.}
   {The cone apex is located at the source region of an eruption instead of the Sun center. The cone axis deviates from the local vertical by an inclination angle of $\theta_1$ and an angle of $\phi_1$. 
    The length and angular width of the cone are $r$ and $\omega$, respectively.}
   {The model is successfully applied to two CMEs originating from the western limb on 2011 August 11 and 2012 December 7.
    By comparing the projections of the cones with the CME fronts simultaneously observed by the Atmospheric Imaging Assembly (AIA) on board the Solar Dynamics Observatory (SDO)
    and the Extreme-Ultraviolet Imager (EUVI) on board the ahead Solar TErrestrial RElations Observatory (STEREO), 
    the properties of the CMEs are derived, including the distance, angular width, inclination angle, deviation from the plane of the sky, and true speed in space.}
   {This revised cone model provides a new and complementary approach in exploring the whole evolutions of CMEs.}


\maketitle

\section{Introduction} \label{s-intro}
Solar prominences are cool and dense plasma suspended in the hot corona \citep{mac10,par14}. 
They usually form through direct injection \citep{chae00}, evaporation-condensation \citep{kar05,hua21}, or levitation-condensation \citep{jen21} processes
along the magnetic polarity inversion lines \citep{vanb89}. The lifetimes of prominences range from a few weeks to a few months.
The destabilization of a prominence may lead to an eruption and evolve into a coronal mass ejection \citep[CME;][]{for00,go00,chen11}.
A typical CME shows a three-part structure: a bright core within a dark cavity surrounded by a bright leading front \citep{ill85}. 
Those originating near the solar disk center and propagating towards the Earth are named frontside halo CMEs \citep{how82,zqm10}.
Considering their potential geoeffectiveness, the prediction of the arrival time of halo CMEs is very important for space weather forecast.
Owing to the strong projection effect, the precise estimation of the space speed (or true speed) becomes difficult. 
Several versions of cone models were proposed based on the assumption that the angular width and velocity of symmetric CMEs keep constant 
in the very early phase \citep[e.g.,][]{how82,zhao02,mich03,xie04,xue05}. Meanwhile, the direction of CMEs is radial and not inclined to the solar normal.
The cone models have been successfully applied to the determination of the space speed, angular width, and arrival times of halo CMEs.
Since a large fraction of CMEs are driven by the eruptions of magnetic flux ropes \citep{che13,yan18}, new models were developed, including the graduated cylindrical shell \citep[GCS;][]{the06} model.

Although solar eruptions are symmetric in most of the theoretical models \citep[e.g.,][]{ant99,chen00,lin00,mo01,jia21}, the eruptions of prominences are not always radial in observations.
Occasionally, the propagations of prominences deviate from the vertical and are termed ``non-radial eruption'' \citep{will05,gos09,shen11,sun12,bi13,kli13,pan13,mc15,yang18,dev21,man21}.
The apparent inclination angles are between $\sim$45$^{\circ}$ and $\sim$90$^{\circ}$. 
Non-radial eruptions are noticed in magnetohydrodynamics (MHD) numerical simulations as well \citep{au10,kli13,jia18}.
In the two-dimensional (2D) simulations of CMEs triggered by emerging flux, the eruption is non-radial when flux emergence occurs away from the neutral line \citep[][see their Fig. 5(b)]{chen00}.

It is noted that the true speeds of non-radial eruptions are still unknown from a single perspective due to the projection effect.
In the cone models mentioned above, the apex is located at the Sun center, which is reasonable and acceptable for radial eruptions. However, this assumption is not applicable for non-radial eruptions.
To determine the geometrical and kinematical properties of CMEs driven by non-radial prominence eruptions, a revised cone model is proposed in this paper. 
The model is applied to two non-radial eruptions simultaneously observed by the Atmospheric Imaging Assembly \citep[AIA;][]{lem12} on board the Solar Dynamics Observatory (SDO) spacecraft
and the Extreme-Ultraviolet Imager (EUVI) in the Sun Earth Connection Coronal and Heliospheric Investigation \citep[SECCHI;][]{how08} package on board the ahead 
Solar TErrestrial RElations Observatory \citep[STEREO;][]{kai08} spacecraft (hereafter STA).
This paper is arranged as follows. The revised cone model is described in Sect.~\ref{s-mod}. The application to the two events is shown in Sect.~\ref{s-app}.
Discussion and a brief summary are given in Sect.~\ref{s-sum}.

\section{Revised cone model} \label{s-mod}
\begin{figure}
\includegraphics[width=8cm,clip=]{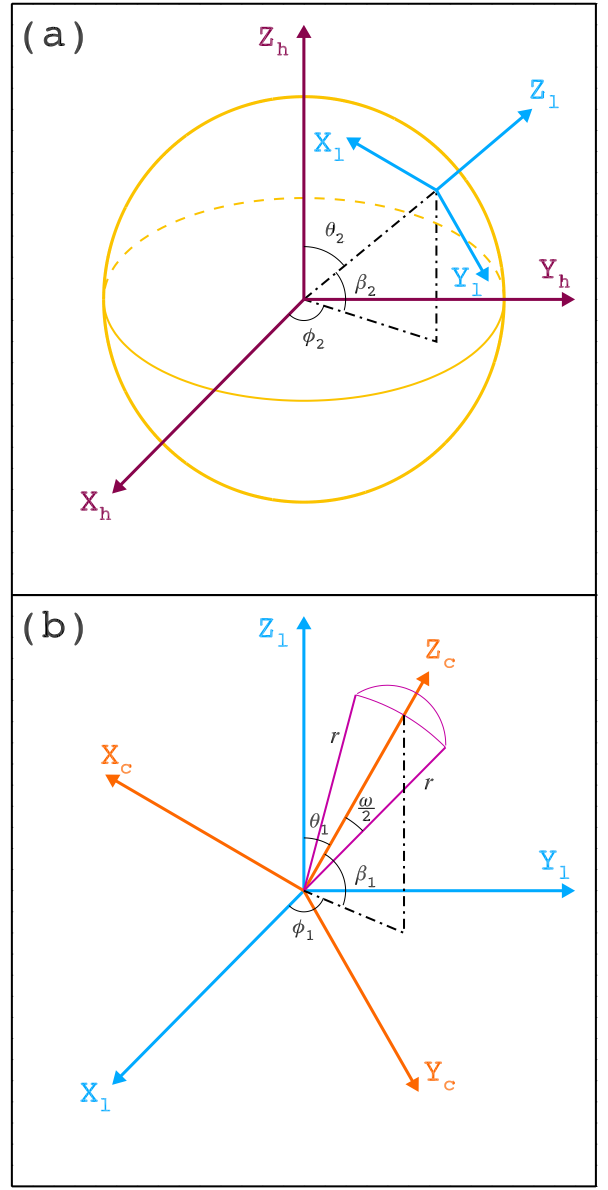}
\centering
\caption{Topology of the revised cone model and the transformations of coordinate systems. See text for details.}
\label{fig1}
\end{figure}

Figure~\ref{fig1} shows the topology of the revised cone model and the transformations of coordinate systems. The heliocentric coordinate system (HCS; $X_h$, $Y_h$, $Z_h$) 
is represented by the dark red axes. The $X_h$ points to the Earth, and ($Y_h$, $Z_h$) defines the plane of the sky (POS).
The local coordinate system (LCS; $X_l$, $Y_l$, $Z_l$) is represented by the sky-blue axes. 
The origin of coordinates denotes the source location of an eruption with a longitude of $\phi_2$ and a latitude of $\beta_2$ ($\theta_2=90^\circ-\beta_2$).
The transform between HCS and LCS is performed by a matrix ($M_2$):

\begin{equation} \label{eqn-1}
\left(
\begin{array}{c}
x_h  \\
y_h  \\
z_h \\
\end{array}
\right)
=M_2
\left(
\begin{array}{c}
x_l  \\
y_l  \\
z_l \\
\end{array}
\right),
\end{equation}
where
\begin{equation} \label{eqn-2}
M_2=
\left(
\begin{array}{ccc}
\cos{\theta_2}\cos{\phi_2} & -\sin{\phi_2} & \cos{\phi_2}\sin{\theta_2}  \\
\cos{\theta_2}\sin{\phi_2}  &  \cos{\phi_2} & \sin{\phi_2}\sin{\theta_2} \\
-\sin{\theta_2}                    &         0          &  \cos{\theta_2} \\
\end{array}
\right).
\end{equation}

In Figure~\ref{fig1}(b), the cone coordinate system (CCS; $X_c$, $Y_c$, $Z_c$) is represented by the orange axes. 
The apex of ice-cream cone is located at the source region and the base of the cone is not planar but spheric \citep{xue05}.
The cone has a length of $r$, an angular width of $\omega$, and its axis is along $Z_c$.
For radial eruption, the CCS is consistent with the LCS. For non-radial eruption, the cone axis deviates from $Z_l$ by angles of $\phi_1$ and $\theta_1$ (i.e., inclination angle).
Likewise, the transform between LCS and CCS is performed by a matrix ($M_1$):

\begin{equation} \label{eqn-3}
\left(
\begin{array}{c}
x_l  \\
y_l  \\
z_l \\
\end{array}
\right)
=M_1
\left(
\begin{array}{c}
x_c  \\
y_c  \\
z_c \\
\end{array}
\right),
\end{equation}
where
\begin{equation} \label{eqn-4}
M_1=
\left(
\begin{array}{ccc}
\cos{\theta_1}\cos{\phi_1} & -\sin{\phi_1} & \cos{\phi_1}\sin{\theta_1} \\
\cos{\theta_1}\sin{\phi_1}  &  \cos{\phi_1} & \sin{\phi_1}\sin{\theta_1} \\
-\sin{\theta_1}                    &         0          &  \cos{\theta_1} \\
\end{array}
\right).
\end{equation}
Therefore, the transform between HCS and CCS is:
\begin{equation} \label{eqn-5}
\left(
\begin{array}{c}
x_h  \\
y_h  \\
z_h \\
\end{array}
\right)
=M_2 M_1
\left(
\begin{array}{c}
x_c  \\
y_c  \\
z_c \\
\end{array}
\right).
\end{equation}

For limb events whose source location is close to the eastern or western limb, the line of sight (LOS) component of velocity could be toward or backward the observer.
Hence, stereoscopic observations from two or more perspectives are required to impose constraints on the geometrical properties of the cone.
Take STEREO for example, the separation angle ($\phi_0$, positive for STA and negative for STB) between the two spacecraft and the Sun-Earth connection increases with time.
The transform between the STEREO coordinate system ($X_e$, $Y_e$, $Z_e$; not shown in Figure~\ref{fig1}) and HCS is performed by a matrix ($M_0$):
\begin{equation} \label{eqn-6}
\left(
\begin{array}{c}
x_e  \\
y_e  \\
z_e \\
\end{array}
\right)
=M_0
\left(
\begin{array}{c}
x_h  \\
y_h  \\
z_h \\
\end{array}
\right),
\end{equation}
where
\begin{equation} \label{eqn-7}
M_0=
\left(
\begin{array}{ccc}
\cos{\phi_0} & \sin{\phi_0} & 0 \\
-\sin{\phi_0} & \cos{\phi_0} & 0 \\
0                  &     0              & 1 \\
\end{array}
\right).
\end{equation}

For non-radial eruptions, the parameters ($\phi_2$, $\theta_2$, $\phi_0$) are known, and the geometrical parameters ($r$, $\omega$, $\phi_1$, $\theta_1$) 
need to be determined according to the simultaneous observations from SDO and STA/STB.

\section{Application to non-radial prominence eruptions} \label{s-app}
\begin{figure}
\includegraphics[width=9cm,clip=]{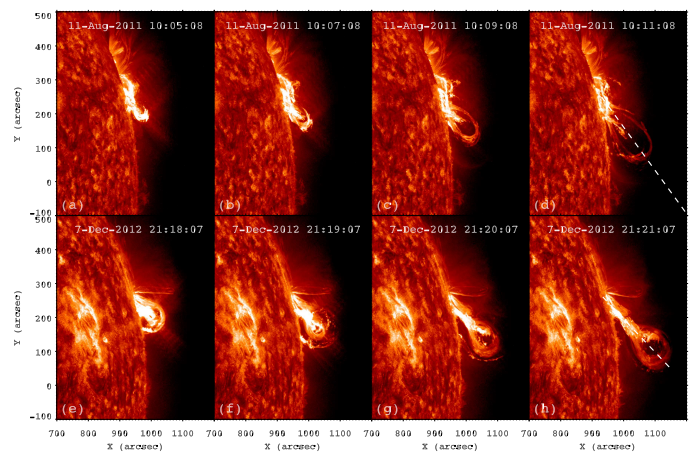}
\centering
\caption{Snapshots of the AIA 304 {\AA} images during the non-radial eruptions on 2011 August 11 (top panels) and 2012 December 7 (bottom panels).
The white dashed lines signify the directions of propagation in the POS.}
\label{fig2}
\end{figure}

To validate the revised cone model, two non-radial eruptions are selected. The first event occurred in NOAA active region (AR) 11263  on 2011 August 11.
\citet{man21} investigated the associated CME and type-{\sc ii} radio bursts. The authors found clear radio evidence for a shock wave reflected by a coronal hole in the south hemisphere (see their Fig. 7). 
In Figure~\ref{fig2}, the top panels show selected AIA 304 {\AA} images during the eruption. It is clear that the loop-like prominence starts to rise before 10:05:00 UT near the western limb 
and propagates in the southwest direction as indicated by the white dashed line. The inclined eruption produced a C6.2 class flare, a fast partial halo CME, and the associated EUV wave \citep{man21}.

\begin{figure*}
\includegraphics[width=14cm,clip=]{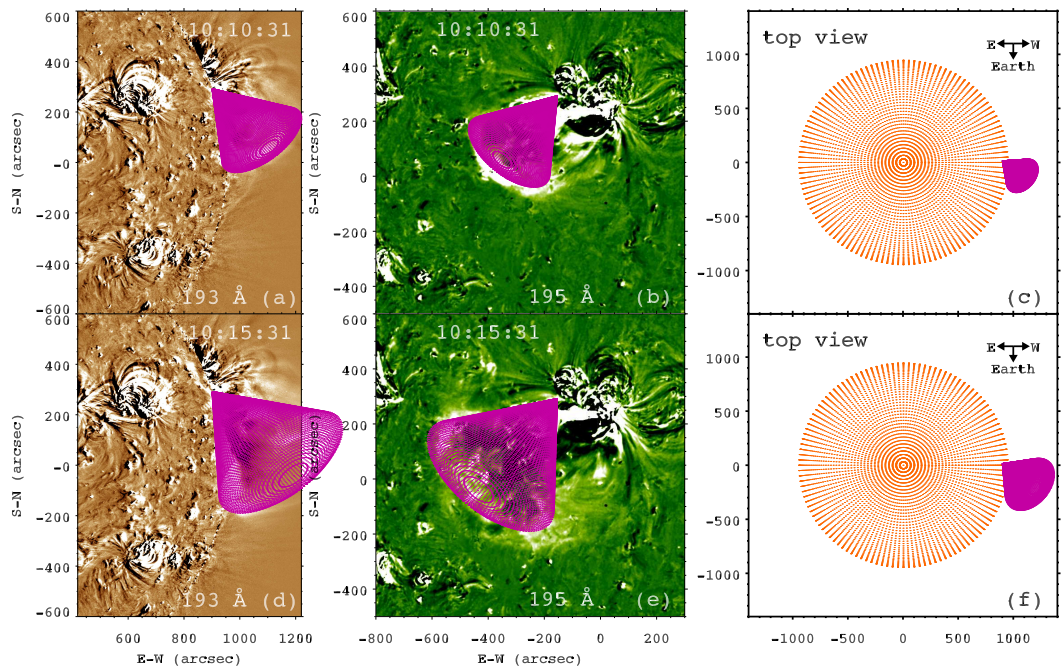}
\centering
\caption{Base-difference images observed by AIA 193 {\AA} (left panels) and STA/EUVI 195 {\AA} (middle panels) at 10:10:31 UT (a-b) and 10:15:31 UT (d-e) on 2011 August 11. 
The projections of the cone are superposed with magenta dots.
Right panels: Top views of the Sun (orange dots) and the associated cones (magenta dots).}
\label{fig3}
\end{figure*}

In Figure~\ref{fig3}, the base-difference images simultaneously observed by AIA 193 {\AA} and STA/EUVI 195 {\AA} at 10:10:31 UT are displayed in panels (a) and (b), respectively.
The bright leading edges represent the EUV wave fronts and also the projection of CME fronts, since the EUV wave front is cospatial with the CME front in the early phase \citep{chen09}.
Assuming that the shape of CME is an ice-cream cone, the projections of the cone in AIA and STA/EUVI field of view are drawn with magenta dots.
It is obvious that the projections of cone fit nicely with the CME front, meaning that the assumption of morphology is also reasonable for non-radial eruptions.
The geometrical parameters of the cone at 10:10:31 UT are listed in the second column of Table~\ref{tab-1}. 
It is seen that the non-radial eruption has an inclination angle ($\theta_1$) of 70$^{\circ}$ from the local vertical. 
A top view of the Sun (orange dots) and the cone (magenta dots) is displayed in Figure~\ref{fig3}(c), showing that the propagation deviates from the POS by 30$^{\circ}$ towards the Earth.
The true speed of CME at 10:10:31 UT is calculated to be $\sim$551 km s$^{-1}$ according to the POS speed ($\sim$477 km s$^{-1}$) and $\phi_1$.
The length ($r$) and angular width ($\omega$) of the cone are 350$\arcsec$ and 60$^{\circ}$. 

Similarly, the base-difference images observed by AIA and STA/EUVI at 10:15:31 UT are displayed in panels (d) and (e), in which the projections of the cone are overlapped with magenta dots.
The front of cone exactly matches the CME front. The parameters are listed in the third column of Table~\ref{tab-1}. Compared with the parameters at 10:10:31 UT, 
the direction of the cone does not change. However, the angular width increases to 66$^{\circ}$, implying a lateral expansion in the initial evolution.
The true speed also increases to $\sim$625 km s$^{-1}$, suggesting an acceleration in the meantime \citep{che13}. It should be emphasized that 
the related CME first appeared in the field of view of Large Angle and Spectroscopic Coronagraph (LASCO) C2 at 10:36:05 UT and propagated in the field of view of LASCO/C3 until $\sim$14:00:00 UT.
The recorded central position angle ($\sim$267$^{\circ}$), angular width ($\sim$167$^{\circ}$), and linear velocity ($\sim$1160 km s$^{-1}$) 
by the CDAW CME catalog \footnote{https://cdaw.gsfc.nasa.gov/CME\_list/} are significantly larger than the derived parameters using the revised model, 
indicating that the revised cone model may not apply in the later phase of a CME.

\begin{figure*}
\includegraphics[width=14cm,clip=]{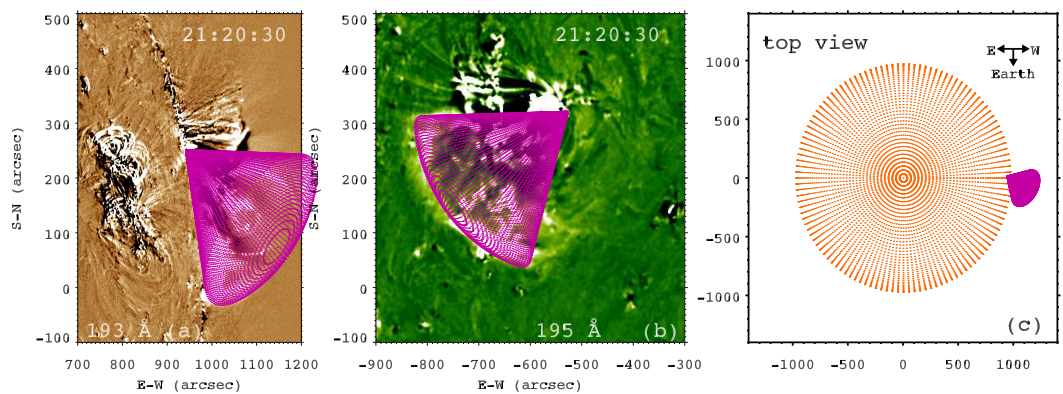}
\centering
\caption{Base-difference images observed by AIA 193 {\AA} (a) and STA/EUVI 195 {\AA} (b) at 21:20:30 UT on 2012 December 7. The projections of the cone are superposed with magenta dots.
(c) Top view of the Sun (orange dots) and the associated cone (magenta dots).}
\label{fig4}
\end{figure*}

The second event occurred in AR 11621 near the western limb on 2012 December 7 when the separation angle between SDO and STA reached 128$^{\circ}$.
The non-radial prominence eruption is illustrated in the bottom panels of Figure~\ref{fig2}. The prominence starts to rise slowly at 21:17:30 UT and rapidly at 21:18:00 UT along the southwest direction.
The inclined eruption also generated a C5.8 class flare, a fast CME, and the associated EUV wave. 
Interestingly, kink oscillations of the adjacent coronal loops were triggered by the eruption, which is the focus of another paper.
In Figure~\ref{fig4}, the base-difference images observed by AIA and STA/EUVI at 21:20:30 UT are displayed in panels (a) and (b), respectively.
The projections of the cone are superposed with magenta dots. Likewise, the front of cone strictly overlaps with the CME front. 
The parameters of the cone are listed in the last column of Table~\ref{tab-1}. The inclination angle $\theta_1$ and $\phi_1$ are 60$^{\circ}$ and -30$^{\circ}$.
A top view of the Sun and cone is displayed in Figure~\ref{fig4}(c), indicating that the CME also deviates from the POS by 30$^{\circ}$ towards the Earth.
The true speed of the CME is calculated to be $\sim$815 km s$^{-1}$ according to the POS speed ($\sim$706 km s$^{-1}$) and $\phi_1$.
The length and angular width of the cone are 300$\arcsec$ and 70$^{\circ}$, respectively.
As mentioned above, the related CME first appeared in the field of view of LASCO/C2 at 21:36:05 UT and propagated until December 8 in the field of view of LASCO/C3 
with a central position angle of $\sim$283$^{\circ}$ and an angular width of $\sim$165$^{\circ}$, which are significantly larger than the values in the early evolution.
Hence, the revised cone model may not apply in the later phase.

\begin{table}
\centering
\caption{Parameters of the two events.} 
\label{tab-1}
\begin{tabular}{lccc}
\hline\hline
Date & 2011-08-11   & 2011-08-11   & 2012-12-07 \\ 
Time &  10:10:31     &  10:15:31      &  21:20:30  \\
\hline
$\phi_0$ ($^{\circ}$) & 101 & 101 & 128 \\
$\phi_2$ ($^{\circ}$) & 91 & 91 & 91 \\
$\theta_2$ ($^{\circ}$) & 72 & 72 & 75 \\
$\beta_2$ ($^{\circ}$) &  18 & 18 & 15 \\
\hline
$\theta_1$ ($^{\circ}$) & 70 & 70 & 60 \\
$\phi_1$ ($^{\circ}$) & -30 & -30 & -30 \\
$r$ (\arcsec) & 350 & 500 & 300 \\
$\omega$ ($^{\circ}$) & 60 & 66 & 70 \\
\hline
$v_{pos}$ (km s$^{-1}$) & 477 & 541 & 706 \\
$v_{true}$ (km s$^{-1}$) & 551 & 625 & 815 \\
\hline
\end{tabular}
\end{table}

\section{Discussion and summary} \label{s-sum}
As mentioned in Sect.~\ref{s-intro}, several versions of cone models were proposed to interpret the geometrical and kinematic properties of halo CMEs \citep{how82,zhao02,mich03,xie04,xue05}.
The traditional cone models assume radial eruptions and the angular width and space speed keep constant during their evolutions. The apex of the cone is located at the Sun center.
However, for non-radial eruptions, the apex should be at the source location of eruption, as indicated in the revised model (Figure~\ref{fig1}).
Moreover, the speed of CME is not necessarily constant (see Table~\ref{tab-1}). For the first event on 2011 August 11, the true speed of CME grows from $\sim$551 to $\sim$625 km s$^{-1}$.
The angular width also increases by a factor of 10\%, suggesting acceleration and lateral expansion. Hence, the revised model is useful in investigating the initial kinematic evolutions of CMEs.
Last but not least, the model has only four free parameters ($\theta_1$, $\phi_1$, $r$, and $\omega$), which is advantageous for a faster 3D reconstruction.

Of course, the revised cone model has limitations. First, it assumes a symmetric morphology like in traditional models, which may not apply to the badly asymmetric events.
Secondly, the model is useful in the very early phase of CMEs before showing up in the white-light coronagraphs, 
since the evolution of a CME has not been strongly influenced by the large-scale, overlying magnetic loops as well as the solar wind \citep{wang14}.
In the later phase of a CME, sophisticated models should be more suitable and complementary \citep{the06,kw14}.
Thirdly, simultaneous observations from two viewpoints are currently used to carry out the fittings. 
Improved results are expected by using stereoscopic observations from three viewpoints (SDO, STA, and STB).
In the future, the revised cone model will hopefully be applied to the CMEs observed by the Lyman-$\alpha$ Solar Telescope \citep[LST;][]{li19}
on board the Advanced Space-based Solar Observatory \citep[ASO-S;][]{gan19} and the Metis \citep{ant20} on board the Solar Orbiter \citep{mu20}.

In this paper, a revised cone model is proposed to investigate the geometrical and kinematic properties of CMEs in their early phases. 
The model is successfully applied to two CMEs as a result of non-radial prominence eruptions originating from the western limb.
By comparing the projections of the cones with the CME fronts simultaneously observed by SDO/AIA and STA/EUVI, 
the properties of the CMEs are derived, including the distance, angular width, inclination angle, deviation from the POS, and true speed in space.
This revised cone model provides an innovative and complementary approach in exploring the complete evolutions of CMEs.

\begin{acknowledgements}
SDO is a mission of NASA\rq{}s Living With a Star Program. AIA data are courtesy of the NASA/SDO science teams.
STEREO/SECCHI data are provided by a consortium of US, UK, Germany, Belgium, and France.
This work is funded by NSFC grants (No. 11773079, 11790302), the International Cooperation and Interchange Program (11961131002), 
and the Strategic Priority Research Program on Space Science, CAS (XDA15052200, XDA15320301).
\end{acknowledgements}

\end{document}